\documentstyle[12pt]{article} 
\newcommand{\bi}{\bibitem}

\newcommand{\Vbf}{\mbox{\boldmath $V$}}
\newcommand{\ubf}{\mbox{\boldmath $u$}}
\newcommand{\imp}{\mbox{\boldmath $p$}}

\newcommand{\Le}{{\cal L}}

\newcommand{\crm}{{\rm c}}

\newcommand{\mrm}{{\rm m}}

\newcommand{\ga}{\gamma}

\newcommand{\infi}{\infty}

\newcommand{\lan}{\langle}

\newcommand{\De}{\Delta}
\newcommand{\ran}{\rangle}
\newcommand{\bb}{\begin{equation}}
\newcommand{\ee}{\end{equation}}
\newcommand{\bega}{\begin{eqnarray}}
\newcommand{\ega}{\end{eqnarray}}
\newcommand{\begae}{\begin{eqnarray*}}
\newcommand{\egae}{\end{eqnarray*}}

\newcommand{\up}{\uparrow}
\newcommand{\dow}{\downarrow}

\newcommand{\h}{\hspace*{4ex}}

\newcommand{\om}{\omega}

\newcommand{\longr}{\longrightarrow}

\newcommand{\ove}{\overline}

\newcommand{\vs}{\vspace*}

\begin{document}

\baselineskip 0.65cm

\begin{center}
{\large {\bf Superluminal motions (and microwave propagation)\\ 
in Special Relativity: Solution of the causal paradoxes.}$^{\: (\dag)}$} 
\footnotetext{$^{\: (\dag)}$ Work partially supported by INFN, MURST and 
CNR, and by CAPES.}
\end{center}

\vspace*{5mm}

\centerline{ Erasmo Recami }

\vspace*{0.5 cm}

\centerline{\em Facolt\`a di Ingegneria,
Universit\`a Statale di Bergamo, Dalmine (BG), Italy;} \centerline{{ 
INFN--Sezione di Milano, Milan, Italy; {\em {\rm and}} }}
\centerline{{\em C.C.S. and D.M.O./FEEC, State University at Campinas, S.P., 
Brazil.}} 

\vspace*{1. cm}

{\bf Abstract  \ --} \ Some recent experiments, performed at Berkeley, 
Cologne, Florence and Vienna led to the claim that something seems to travel 
with a speed larger than the speed $c$ of light in vacuum. Various other 
experimental results seem to point in the same direction: For instance, 
localized wavelet-type solutions of Maxwell equations have been found, both 
theoretically and experimentally, that travel with Superluminal speed. Even 
muonic and electronic neutrinos  ---it has been proposed---  might be 
``tachyons", since their square mass 
appears to be negative; not to mention the apparent Superluminal expansions 
observed in the core of quasars and, recently, in the so-called galactic
microquasars. \ In the first part of this paper we verify, on the basis 
of the numerical solution of Maxwell equations, that waves propagating down
a microwave guide can travel with Superluminal group velocity, just confirming
some of the previously mentioned experimental results. \ Then, we have to 
face the question of Superluminal motions within the theory of Special 
Relativity. \ It is not widely recognized that all such theoretical and
experimental results do {\em not} place relativistic causality in jeopardy. \ 
For instance, it is possible (at least in microphysics) to solve also the 
known causal paradoxes, devised for ``faster than light" motion. \ 
Here we show, in detail and rigorously, how to solve the oldest causal 
paradox, originally proposed by Tolman, which is the kernel of many further 
tachyon paradoxes.  The key to the solution is a careful application of 
tachyon mechanics, as it unambiguously follows from Special Relativity.\\

\

{\bf Introduction.} --  Superluminal propagation has been observed 
in several areas of physics($^1$). \
In electromagnetism, in particular, some recent experiments performed at 
Cologne($^2$), Berkeley($^3$), Florence and Vienna with
evanescent waves (``tunnelling photons") led to the claim that evanescent 
modes can travel with a group velocity larger than the speed 
$c$ of light in vacuum, thus confirming some older predictions($^4$). \
Even more recently, some of the main experimental claims (e.g., in ($^2$)) 
have been verified($^5$) just by solving the Maxwell equations with the 
requested boundary conditions.

\h Various other 
experimental results seem to point in the same direction: For instance, 
localized wavelet-type solutions of Maxwell equations have been found, both 
theoretically($^1$) and experimentally($^6$), that travel with Superluminal 
speed. \ Even muonic and electronic neutrinos  ---it has been proposed---  
might be ``tachyons", since their square mass appears to be negative($^7$); 
not to mention the apparent Superluminal expansions observed in the core of 
quasars($^8$) and, recently, in the so-called galactic microquasars($^9$). 

\ Nevertheless, all such data or results do {\em not} seem to place 
relativistic causality in jeopardy. \ In fact, when Special Relativity (SR)
is not restricted to subluminal objects, so that it expands into the 
so-called Extended Relativity, one is left with a theory describing
also Superluminal objects and waves on the basis of the ordinary postulates
of SR. As far as the foundations of Extended Relativity are concerned,
we shall only quote ref.($^{10}$) and references therein. \  Here we shall
mainly confine our attention to the fact that it is possible to solve (also) 
the known causal paradoxes, devised for ``faster than light" motion, even if 
this is not widely recognized.

\h We shall come back soon to this point. \ First, however, let us briefly
examine for instance the experiments in refs.($^2$), by making recourse to 
work performed by A.Pablo L.Barbero and H.E. Hern\'andez-Figueroa at 
UNICAMP.($^{5}$)\\

{\bf Theoretical verification of microwave Superluminal propagation.} --  The 
inset (a) in Fig.1 shows the waveguide used in those experiments.
Launching a TE$_{10}$ signal, limited in frequency by $\om_\mrm$ such that
$\om_\mrm < \om_\crm$, where $\om_\crm$ is the cutoff frequency for the
fundamental mode in the ``undersized" waveguide (or ``barrier"), those 
authors showed that the transit time of the signal along the barrier is 
independent of its length $a$. \ This agrees, incidentally, with the effect 
predicted by quantum mechanics for particle tunnelling($^{11}$): an effect that 
we called ``Hartman effect". Considering that purely evanescent waves travel
in principle with infinite speed, the finite Superluminal speed encountered 
in those measurements can be attributed to the delay induced by the 
geometrical discontinuities present at the barrier edges. Using a spectrum
analyzer, the transfer function associated with the barrier can be gotten
by measuring the $S$-parameter. The phase of the transfer function for four
different barrier lengths (40, 60, 80 1nd 100 mm) is shown in Fig.2(b); notice
that such four curves are practically superposed.($^{5}$)

\h In order to check those experimental results, in ref.($^{5}$) it was 
adopted a numerical code 
based on the method of moments (MoM), widely used in the design of microwave
filters and other devices by the Antennas Research Group at CPqD-Telebr\'as,
Brazil.  Such a code solves the Maxwell equations in the frequency domain, 
in presence of transitions associated with inductive (vertical) 
discontinuities only (Fig.1, inset $b$). \ Notice that the experimental
setup in inset $a$ of Fig.1, is not really very different, even if it takes
into account also capacitive (horizontal) discontinuities: the latter, in
fact, produce a transfer function whose phase, as a function of the frequency, 
is practically constant, and their magnitude is much smaller than the one 
produced by the inductive discontinuities.($^{12}$)

\h We adopted($^{5}$) a pulse modulated by a Kaiser-Bessel window, 
producing a signal with a limited spectrum, between 8.2 and 9.2 GHz, similar 
to the input signal used in ($^2$).  The transfer functions were computed 
for barriers of 40, 60 and
100 mm.  The phases of such functions are shown in Fig.2a: notice their 
superposition. The maximum magnitudes of the output pulses were 0.05743, 
0.01248 and 0.00062 for the 40, 60 and 100 mm barriers, respectively.  These
values were normalized with respect to the maximum magnitude of the input
pulse, and exhibit an error of less than 4\% relative to the experimental
results($^2$). \ The computed phase (Fig.2a) and the measured one (Fig.2b)
are in qualitative agreement; they differ by a shift of about 10 degrees, 
due ---as already mentioned--- to the capacity discontinuities, considered in
the experiment and ignored in our simulations. \ Next, by using the Fourier 
transform, we computed($^{5}$) the pulse propagation in the time domain: 
the transit
time for the input signal to travel down the three different barriers (of 40,
60 and 100 mm) was {\em the same\/}: 115 ps. \ Figure 1 shows the output
signals for the three barriers. \ In refs.($^2$) the measured transit time was 
130 ps; the difference of 15 ps may be due again to the capacity 
discontinuities.

\h In conclusion, just by solving the Maxwell equations, we can confirm 
the main experimental results in ($^2$). \ In ref.($^{13}$), experiments with two 
barriers suggest the possibility of long-distance Superluminal propagation:
we shall examine that case elsewhere($^{5}$). \  Here we want to discuss a more
general question. If something can travel with a speed larger than the speed 
$c$ of light in vacuum, as suggested not only by the previous considerations
but also by the results known from the abovementioned three further sectors 
of experimental physics (see Appendix B), then one has to face the problem of 
Superluminal motions from the point of view of Special Relativity. \ We have
already claimed that all the previous results do not seem to place 
relativistic causality in jeopardy. \ In fact, when Special Relativity (SR)
is not restricted to subluminal objects, so that it expands into the 
so-called Extended Relativity, one is left with a theory describing
also Superluminal objects and waves on the basis of the ordinary postulates
of SR. As far as the foundations of Extended Relativity are concerned,
we shall only quote ref.($^{14-16}$) and references therein, and in particular
ref.($^{10}$); while we shall expound what is necessary for our present
purposes in the Sections below and in Appendix A.

\h Let us then pass to the problem of solving the causal paradoxes arising for
for ``faster than light" motions.\\

{\bf Superluminal motions and relativistic causality} --  In fact, claims 
exist since long that all the ordinary causal paradoxes 
proposed for tachyons can be solved($^{14-16}$) (at least ``in microphysics")
on the basis of the ``switching procedure" (swp) introduced 
by St\"{u}ckelberg($^{17}$), Feynman($^{17}$) and Sudarshan($^{14}$), also
known as the reinterpretation principle: a principle which in refs.($^{10,15}$)
by Recami et al. has been given the status of a fundamental postulate 
of special relativity, both for bradyons [slower--than--light
particles] and for tachyons. \ Schwartz,($^{18}$) at last, gave the swp
a formalization in which it becomes ``automatic".

\h However, the effectiveness of the swp and of that solution is often 
overlooked, or misunderstood. \ 
Here we want therefore to show, in detail and rigorously,
how to solve the oldest ``paradox", i.e. the {\em antitelephone} one,
originally proposed by Tolman($^{19}$) and then reproposed by many
authors. We shall refer to its recent formulation by Regge,($^{20}$)
and spend some care in solving it, since it is the kernel of many other 
paradoxes.  Let us stress that: \ (i) any careful solution of the tachyon 
causal ``paradoxes" has to make recourse to explicit calculations based on 
the mechanics of tachyons; \ (ii) such tachyon mechanics can be
unambiguously and uniquely derived from SR, by referring the
Superluminal ($V^2 > c^2$) objects to the class of the ordinary, subluminal 
($u^2 < c^2$)  observers
{\em only} (i.e., without any need of introducing ``Superluminal
reference frames''); \ (iii) moreover, the comprehension of the whole 
subject will be substantially enhanced if one refers himself to
the (subluminal, ordinary) SR based on the {\em whole} proper Lorentz group 
$\Le_+ \equiv \Le^\up_+ \cup
\Le^\dow_+$, rather than on its orthochronous subgroup $\Le^\up_+$ only
[see refs.($^{21}$), and references therein]. \ At last, for a modern
approach to the classical theory of tachyons, reference can be made to 
the review article($^{10}$) as well as to refs.($^{15,16}$).\\

\h Before going on, let us mention the following. It is a known fact that 
in the time-independent case the (relativistic, non-quantistic) Helmholtz 
equation and the (non-relativistic, quantum) Schroedinger equation are 
formally identical($^{22}$) [in the time-dependent case, such equations become 
actually different, but nevertheless strict relations still hold between 
some solutions of theirs, as it will be explicitly shown elsewhere($^{23}$)]: 
one 
important consequence of this fact being that evanescent wave transmission 
simulates electron tunnelling. On the other side, a wave-packet had been
predicted($^{24}$) since long to tunnel through an (opaque) barrier with 
Superluminal group-velocity. Therefore, one could expect evanescent waves
too to be endowed with Superluminal (group) speeds.($^{25}$) The abovenamed 
experiments($^{1-3}$), which seem to have actually verified such an 
expectation, are the ones that most attracted the attention of the scientific
press.($^{26}$) \ But they are not the only ones which seem to indicate the 
existence of Superluminal motions.$^{27}$ \\  

\h {\em Tachyon mechanics.} -- In refs.($^{28}$) the basic tachyon mechanics
can be found exploited for the processes: \ a) proper (or
``intrinsic") emission of a tachyon T by an ordinary body A; \ b)
``intrinsic" absorption of a tachyon T by an ordinary body A; where the term
``intrinsic" refers to the fact that those processes
(emission, absorption by A) are described {\em as they appear} 
in the rest-frame of A; particle T can represent both a tachyon and an 
antitachyon. Let us recall the following results only.

\h Let us first consider a tachyon moving with velocity $\Vbf$ in
a reference frame $s_0$. If we pass to a second frame $s'$, endowed with
velocity $\ubf$ w.r.t. (with respect to) frame $s_0$, then the new
observer $s'$ will see ---instead of the initial tachyon T--- an
antitachyon $\ove{\rm T}$ travelling the opposite way in space (due to the
swp: cf. Appendix A), if and only if\\

(1) {\hfill{
$ \ubf \cdot \Vbf > c^2 \; . $
\hfill}} 

\

Recall in particular that, if $\ubf \cdot \Vbf < 0$, the ``switching"
does {\em never} come into play.

\h Now, let us explore some of the unusual and unexpected
consequences of the trivial fact that in the case of tachyons it is\\

(2) {\hfill{
$ |E| = + \sqrt{\imp^2-m^2_0} \qquad (m_0 \ \ \mbox{real}; \ {\Vbf}^2 > 1) 
\; , $
\hfill}} 

\

where we chose units so that, numerically, $\: c=1$.

\h Let us, e.g., describe the phenomenon of ``intrinsic emission" of
a tachyon, as seen in the rest frame of the emitting body: Namely,
let us consider {\em in its rest frame} an ordinary body A, with
initial rest mass $M$, which emits a tachyon (or antitachyon) T
endowed with (real) rest mass($^6$) $m \equiv m_0$, four-momentum $p^\mu
\equiv (E_{\rm T}, \imp)$, and velocity $\Vbf$ along the $x$-axis. 
Let $M'$ be the
final rest mass of body A. The four-momentum conservation requires\\

(3) {\hfill{
$M= \sqrt{\imp^2-m^2} + \sqrt{\imp^2+ M'^{\, 2}}$
\hfill}} (rest frame)\\

that is to say [$V \equiv |\Vbf|$]:\\

(4) {\hfill{
$ 2M |\imp| = [(m^2+ \De)^2 + 4m^2 M^2]^{\frac{1}{2}} \ ; \quad V= [1+
4m^2 M^2/ (m+\De)^2]^{\frac{1}{2}} \; , $
\hfill}}

\

where [calling  $E_{\rm T} \equiv + \sqrt{\imp^2-m^2} \,$]:\\

(5) {\hfill{
$\De\equiv M'^{\, 2} - M^2= -m^2- 2ME_{\rm T} \; ,$ 
\hfill}} (emission)\\

so that\\

(6) {\hfill{
$-M^2 < \De \leq - |\imp|^2 \leq - m^2 \; .$
\hfill}} (emission)\\

It is essential to notice that $\De$ is, of course, an {\em
invariant} quantity, which in a generic frame $s$ writes\\

(7) {\hfill{
$ \De= -m^2 - 2p_\mu P^\mu \; , $
\hfill}}

\

where $P^\mu$ is the initial four-momentum of body A w.r.t. frame
$s$.

\h Notice that in the generic frame $s$ the process of (intrinsic)
emission can appear either as a T emission or as a $\ove{\rm T}$ absorption
(due to a possible ``switching") by body A. \ The following theorem, however,
holds:($^{28}$)\\

Theorem 1: \ $<<$ A necessary and sufficient condition for a
process to be a tachyon emission in the A rest-frame (i.e., to be
an {\em intrinsic emission}) is that during the process the body A
{\em lowers} its rest-mass (invariant statement!) in such a way that 
$-M^2 < \De \leq -m^2$.~$>>$\\

\h Let us now describe the process of ``intrinsic absorption" of a
tachyon by body A; \ i.e., let us consider an ordinary body A to
absorb {\em in its rest frame} a tachyon (or antitachyon) T,
travelling again with speed $V$ along the $x$-direction. The
four-momentum conservation now requires\\

(8) {\hfill{
$M+ \sqrt{\imp^2-m^2} = \sqrt{\imp^2+ M'^{\, 2}} \; ,$
\hfill}} (rest frame)\\

which corresponds to\\

(9) {\hfill{
$\De \equiv M'^{\, 2} - M^2= -m^2+ 2 M E_{\rm T}  \; ,$
\hfill}} (absorption)\\

so that\\

(10) {\hfill{
$-m^2 \leq \De \leq + \infi \; .$
\hfill}} (absorption)\\

In a generic frame $s$, the quantity $\De$ takes the invariant form\\

(11) {\hfill{
$ \De= -m^2+ 2 p_\mu P^\mu \; . $
\hfill}}

\

It results in the following new theorem:\\

Theorem 2: \ $<<$ A necessary and sufficient condition for a
process (observed either as the emission or as the absorption of a
tachyon T by an ordinary body A) to be a tachyon absorption in
the A-rest-frame  ---i.e., to be an {\em intrinsic absorption}---  is
that $\De \geq - m^2$.~$>>$\\

We now have to describe the {\em tachyon exchange} between two
ordinary bodies A and B. We have to consider the four-momentum
conservation at A {\em and} at B; we need to choose a (single) frame
relative to which we describe the whole interaction; let us choose the
rest-frame of A. Let us explicitly remark, {\em however}, that  ---when
bodies A and B exchange one tachyon T---  the tachyon mechanics
is such that the ``intrinsic descriptions" of the processes at A
{\em and} at B can a priori correspond to one of the following four
cases($^{28}$):

\

\setcounter{equation}{11}
\bb
\left\{\begin{array}{ll}
1) & \quad\mbox{emission---absorption} \ ,\\
\\
2) & \quad\mbox{absorption---emission} \ ,\\
\\
3) & \quad\mbox{emission---emission} \ ,\\
\\
4) & \quad\mbox{absorption---absorption} \ .
\end{array}\right.
\ee

\

Case 3) can happen, of course, only when the tachyon exchange takes
place in the receding phase (i.e., while A, B are receding from
each other); case 4) can happen, by contrast, only in the
approaching phase.

\h Let us consider here only the particular tachyon exchanges in
which we have an ``intrinsic emission" at A, and in which moreover the
velocities $\ubf$ of B and $\Vbf$ of T w.r.t. body A are such that $\ubf
\cdot \Vbf > 1$. \ Because of the last condition and the consequent
``switching" (cf. Eq.(1)), from the rest-frame of B one will therefore
observe the flight of an antitachyon $\ove{\rm T}$ {\em emitted} by B and 
absorbed by A \ (the {\em necessary} condition for this to happen, let us 
recall, being  that A, B  {\em recede} from each other).

\h More generally, the dynamical conditions for a tachyon to be
exchangeable between A and B can be shown to be the
following:\\

I) \ Case of ``intrinsic emission" at A:
\bb
\left\{\begin{array}{l}
 \ \mbox{if} \ \ubf \cdot \Vbf < 1 \ , \quad\mbox{then} \ \De_{\rm B} > - 
m^2 \quad (\longr \mbox{intrinsic absorption at B});\\
\\
 \ \mbox{if} \ \ubf \cdot \Vbf > 1 \ , \quad\mbox{then} \ \De_{\rm B} < - 
m^2 \quad (\longr \mbox{intrinsic emission at B}).\\
\end{array}\right.
\ee

II) \ Case of ``intrinsic absorption" at A:
\bb
\left\{\begin{array}{l}
 \ \mbox{if} \ \ubf \cdot \Vbf < 1 \ , \quad\mbox{then} \ \De_{\rm B} < - 
m^2 \quad (\longr \mbox{intrinsic emission at B});\\
\\
 \ \mbox{if} \ \ubf \cdot \Vbf > 1 \ , \quad\mbox{then} \ \De_{\rm B} > - 
m^2 \quad (\longr \mbox{intrinsic absorption at B}).\\
\end{array}\right.
\ee

\h Now, let us finally pass to examine the Tolman paradox.\\

\h {\em The paradox.} -- In Figs.3, 4 the axes $t$ and $t'$ are the
world-lines of two devices A and B, respectively, which are able to
exchange tachyons and move with constant relative speed $u$, [$u^2 <
1$], along the $x$-axis. According to the terms of the paradox (Fig.3),
device A sends tachyon 1 to B (in other words, tachyon 1 is
supposed to move forward in time w.r.t. device A). The device B
is constructed so as to send back tachyon 2 to A as soon as it
receives tachyon 1 from A. If B has to {\em emit} (in its
rest-frame) tachyon 2, then 2 must move forward in time w.r.t. device
B; that is to say, the world-line ${\rm BA}_2$  must have a slope {\em
lower} than the slope ${\rm BA}'$ of the $x'$-axis (where ${\rm BA}' // x'$): 
 \ this means that ${\rm A}_2$ must stay above ${\rm A}'$. If the speed of 
tachyon 2 is
such that ${\rm A}_2$ falls between ${\rm A}'$ and ${\rm A}_1$, it seems 
that 2 reaches A (event ${\rm A}_2$) {\em before} the emission of 1 
(event ${\rm A}_1$).
This appears to realize an {\em anti-telephone}.\\

\h {\em The solution.} -- First of all, since tachyon 2 moves
backwards in time w.r.t. body A, the event ${\rm A}_2$ will appear to
A as the emission of an antitachyon $\ove{2}$. \ The observer ``$\: t \:$"
will see his own device A (able to exchange tachyons) emit
successively towards B the antitachyon $\ove{2}$ and the tachyon 1.

\h At this point, some supporters of the paradox (overlooking tachyon
mechanics, as well as relations (12)) would say that, well, the
description put forth by the observer ``$\: t \:$" can be orthodox, but
then the device B is no longer working according to the stated programme,
because B is no longer emitting a tachyon 2 on receipt of tachyon
1. \ Such a statement would be wrong, however, since the fact that
``$\: t \:$" observes an ``intrinsic emission" at ${\rm A}_2$ {\em does not 
mean} that ``$\: t' \:$" will see an ``intrinsic absorption" at B! \ On the
contrary, we are just in the case considered above, between eqs. (12)
and (13): intrinsic emission by A, at ${\rm A}_2$, with $\ubf \cdot
\Vbf_{\ove{2}} > c^2$, where $\ubf$ and $\Vbf_{\ove{2}}$ are the velocities of
B and $\ove{2}$ w.r.t. body A, respectively; so that {\em both}
A {\em and} B experience an intrinsic {\em emission} (of tachyon 2 or of
antitachyon $\ove{2}$) in their own rest frame.

\h But the tacit premises underlying the  ``paradox" (and even the very
terms in which it was formulated) were ``cheating" us 
{\em ab initio}. \ In fact, Fig.3 makes it clear that, if $\ubf \cdot
\Vbf_{\ove{2}} > c^2$, then for tachyon 1 {\em a fortiori} $\ubf \cdot \Vbf_1 >
c^2$, where $\ubf$ and $\Vbf_1$ are the velocities of B and 1 w.r.t. body
A. \ Therefore, due to the previous consequences of tachyon mechanics, 
observer
``$t' \,$" will see B intrinsically {\em emit} also tachyon 1 (or, rather,
antitachyon $\ove{1}$). \ In conclusion, the proposed chain of events does
{\em not} include any tachyon absorption by B (in its rest frame).

\h For body B to {\em absorb} (in its own rest frame) tachyon 1,
the world-line of 1 ought to have a slope {\em higher} than the slope
of the $x'$-axis (see Fig.4). Moreover, for body B to {\em emit}
(``intrinsically") tachyon 2, the slope of the of 2 should be lower
than the $x'$-axis'. In other words, when the body B, programmed to
emit 2 as soon as it receives 1, does actually do so, the event ${\rm A}_2$
does happen {\em after} ${\rm A}_1$ (cf. Fig.4), as requested by 
causality.\\

\h {\em The moral.} -- The moral of the story is twofold: \ i) one
should never {\em mix} the descriptions (of one phenomenon)
yielded by different observers; otherwise ---even in ordinary 
physics--- one would  immediately meet contradictions: in Fig.3, e.g., the
motion direction of 1 is assigned by A and the motion-direction of
2 is assigned by B; this is ``illegal"; \ ii) when proposing a problem
about tachyons, one must comply($^{14}$) with the rules of tachyon
mechanics($^{25}$); this is analogous to complying with the laws of
{\em ordinary} physics when formulating the text of an {\em ordinary}
problem (otherwise
the problem in itself will be ``wrong").

\h Most of the paradoxes proposed in the literature suffered the
above shortcomings.

\h Notice once more that, in the case of Fig.3, neither A nor B regard
event ${\rm A}_1$ as the cause of event ${\rm A}_2$ (or {\em vice-versa}). 
In the case of Fig.4, on the other hand, both A and B consider event
${\rm A}_1$ to be the cause of event ${\rm A}_2$: but in this case 
${\rm A}_1$ does
chronologically precede ${\rm A}_2$ according to both observers, in
agreement with the relativistic covariance of the law of retarded
causality.\hfill\break

\vs{3. cm}  

\h The author is grateful for stimulating discussions or generous
collaboration to A.P.L.Barbero, R.Bonifacio, G.Degli Antoni, F.Fontana,
F.Hehl, H.E.Hern\'andez F., L.C.Kretly, G.La Pietra, T.Mietzner G.Nimtz, 
L.Salvo, 
C.Sedda, D.Stauffer, J.W.Swart, G.Tarozzi, and gladly acknowledges stimulating 
discussions with Y.Akebo, L.Bosi, GC.Cavalleri, R.Chiao, R.Colombi, S.Giani, 
A.Gigli Berzolari, M.Jammer, D.Jwroszynshi, G.Kurizki, J.Le\'on, 
P.-O.L\"{o}wdin, 
J.-y.Lu, R.Mignani, G.Salesi, A.Steinberg, E.C.G.Sudarshan, C.Ussami 
and Sir Denys Wilkinson. \ He thanks also Professor U.Gerlach for a careful 
reading of an early version of this manuscript.\hfill\break

\vfill\eject

\vs{1 cm} 

{\bf APPENDIX A:} The Stueckelber--Feynman--Sudarshan ``Switching Principle" \\

\h What follows refers equally well to bradyons and to tachyons. For
simplicity, then, let us fix our attention in this Appendix only to the 
case of bradyons. Let us start from a positive-energy particle P travelling
forward in time. As well-known, any orthochronows LT, \ $L^\uparrow$, \
transforms it into another particle still endowed with positive
energy and motion forward in time. On the contrary, any antichronous
(=non-orthochronous) LT, \ $L^\downarrow = - L^\uparrow$, \ will change 
sign --among the others--
to the time-components of {\em all the four-vectors} associated with
P. Any $L^\downarrow$ will transform P into a particle P'
endowed in particular with negative energy {\em and} motion backwards
in time (Fig.5). \ We are of course assuming that $<<$negative-energy 
objects travelling forward in time do {\em not} exist$>>$. (Elsewhere
this Assumption has been given by us the status of a fundamental 
postulate). 

\h In other words, SR together with the natural Assumption above, {\em
implies} that a particle going backwards in time (G\"{o}del($^{29}$))
(Fig.5) corresponds in the four-momentum space, Fig.6, to a particle
carrying negative energy; and, vice-versa, that changing the energy
sign in the latter space corresponds to changing the sign of time in the 
former (dual) space. It is then easy to see that these two paradoxical 
occurrences (``negative energy" and ``motion backwards in time") give rise 
to a phenomenon  that
any observer will describe in a quite {\em orthodox} way, when they
are --as they actually are-- simultaneous (Recami($^{10,14-16}$) and
refs. therein). 

\h Notice, namely, that: \ (i) every observer (a macro-object) explores
space-time, Fig.5, in the positive $t$-direction, so that we shall
meet $B$ as the first and $A$ as the last event; \ (ii) emission
of positive quantity is equivalent to absorption of negative
quantity, as $(-) \cdot (-) = (+) \cdot (+)$; and so on. 

\h Let us now suppose (Fig.7) that a particle P' with negative energy
(and, e.g., charge $-e$), travelling backwards in time, is emitted by
A at time $t_1$ and absorbed by B at time $t_2 < t_1$. Then, it
follows that at time $t_1$ the object A ``looses" negative energy and negative
charge, i.e. {\em gains} positive energy and positive charge. And that at
time $t_2 < t_1$ the object B ``gains" negative energy and charge,
i.e. {\em looses} positive energy and charge. The physical phenomenon
here described is nothing but the exchange {\em from} B {\em to}
A of a particle Q with {\em positive} energy, charge $+e$, and travelling 
{\em forward} in time. Notice that Q has,
however, charges {\em opposite} to P'; this means that
the present ``switching procedure" (previously called also ``RIP") effects
a ``charge conjugation" $C$, among
the others. Notice also that ``charge", here and in the following,
means {\em any} additive charge; so that our definitions of charge
conjugation, etc., are more general than the ordinary ones (Recami
and Mignani,($^{30}$) hereafter called Review I; Recami($^{10,15}$)).
Incidentally, such a switching procedure has been shown to be
equivalent to applying the chirality operation $\ga_5$ (Recami and
Ziino($^{31}$)).\\

{\em Matter and Antimatter from SR} --  A close inspection shows that the 
application of any antichronous transformation $L^\downarrow$, together with 
the switching procedure, transforms P into an object

\bb
{\rm Q} \equiv \ove{\rm P} 
\ee
which is indeed the {\em antiparticle} of P. We are saying that {\em
the concept of antimatter is a purely relativistic one}, and that, on
the basis of the double sign in $[c=1]$ 

\bb
E = \pm \sqrt{\imp^2 + m^2_0} ,
\ee
the existence of antiparticles could have been predicted already in 1905,
exactly with the properties they actually exhibited when later
discovered, {\em provided that} recourse to the ``switching procedure" had
been made. We therefore maintain that the points of the lower
hyperboloid sheet in Fig.6 --since they corresponds not only to
negative energy but also to motion backwards in time-- represent
the kinematical states of the {\em antiparticle} $\ove{\rm P}$ {\em of}
the particle P represented by the upper hyperboloid sheet). 

\h Let us stress that the switching procedure not only can, but {\em
must} be enforced, since any observer can do nothing but explore
spacetime along the positive time direction. That procedure is an improved
translation into a purely relativistic language of the
St\"{u}ckelberg--Feynman($^{17}$) ``Switching principle". Together with
our  Assumption above, it can take the form of a ``Third Postulate":
$\lan\lan$Negative-energy objects travelling forward in time do {\em not}
exist; any negative-energy object P travelling backwards in time
can and must be described as its antiobject $\ove{\rm P}$ going the
opposite way {\em in space} (but endowed with positive energy and motion
forward in time)$\ran\ran$. Cf. e.g. Caldirola and Recami($^{32}$),
Recami($^{10,15}$) and references therein.\\

{\em Concluding remark of Appendix A} --  Let us go back to Fig.5. In SR, when based only 
on the two ordinary postulates, nothing prevents a priori the event $A$ from
influencing the event $B$. Just to forbid such a possibility we
introduced our Assumption together with the
Switching procedure. As a consequence, not only we eliminate any
particle-motion backwards in time, but we also ``predict" and
naturally explain within SR the existence of antimatter.

\h In the case of tachyons the Switching procedure was first applied by
Sudarshan and coworkers($^{14}$);  see e.g. ref.($^{15}$) and refs. therein.\hfill\break

\vfill\eject

\vs{1 cm} 

{\bf APPENDIX B:} Experimental data possibly related to Superluminal motions.  \\

\h The question of Super-luminal objects or waves
[tachyons: a term coined by G.Feinberg] has a long story, starting perhaps
with Lucretius' {\em De Rerum Natura} (cf., e.g., book 4, line 201). Still
in pre-relativistic times, let us recall e.g. the papers by A.Sommerfeld
(quoted in refs.($^{10,33}$)). In relativistic times, our problem started to be
tackled again essentially in the fifties and sixties, in particular after
the papers by E.C.George Sudarshan et al., and later on by E.Recami,
R.Mignani, et al. [who coined the term bradyons for slower-than-light
objects, and rendered the expressions subluminal and Superluminal of popular
use by their works at the beginning of the seventies], as well as by
H.C.Corben and others (to confine ourselves to the {\em theoretical}
researches). For references, one can check pages 162-178 in ref.($^{10}$), where
about 600 citations are listed; as well as the large bibliographies by
V.F.Perepelitsa($^{34}$) and as the book in ref.($^{16}$). \ In particular, for 
the causality problems one can see refs.($^{10,15}$) and references therein, 
while for a model theory for tachyons in two dimensions one can be addressed 
to refs.($^{10,17}$). \ The first experiments looking for tachyons were 
performed by
T.Alv\"{a}ger et al.; some citations about the early experimental quest for
Superluminal objects may be found e.g. in refs.($^{35}$).

\h The subject of tachyons is presently returning after fashion,
especially because of the fact that four different experimental
sectors of physics {\em seem} to indicate the existence of Superluminal 
objects. One of such sectors has been already mentioned by us. We wish to 
put forth in the following some information (mainly
bibliographical) about the experimental results obtained in the other
sectors.\\

{\em Second - Negative Mass-Square Neutrinos} -- Since 1971 it was known 
that the experimental square-mass of {\em muon}-neutrinos resulted to be 
negative (with low statistical significance, but systematically). If 
confirmed, this would correspond (within the ordinary, na\"{\i}ve approach to 
relativistic particles) to an imaginary mass and therefore to a Superluminal 
speed; in a non-na\"{\i}ve approach,($^{10}$)
i.e. within a Special Relativity theory extended to include tachyons
[Extended Relativity (ER)], the free tachyon ``dispersion relation" becomes
\ $E^2 - {\mbox{\boldmath $p$}}^2 = -m_{{\rm o}}^2$. \ See e.g. \ E.V.Shrum
and K.O.H.Ziock: Phys. Lett. B37 (1971) 114; \ D.C.Lu et al.: Phys. Rev.
Lett. 45 (1980) 1066; \ G.Backenstoss et al.: Phys. Lett. B43 (1973) 539; \
H.B.Anderhub et al.: Phys. Lett. B114 (1982) 76; \ R.Abela et al.: Phys.
Lett. B146 (1984) 431;  \ B.Jeckelmann et al.: Phys. Rev. Lett. 56 (1986)
1444.

\h From the theoretical point of view, about the above point see
E.Giannetto, G.D. Maccarrone, R.Mignani and E.Recami: Phys. Lett. B178 (1986)
115-120, and references therein; \ see also ref.$^{10}$ and references 
therein.

\h Recent experiments showed that also {\em electron}-neutrinos result
to have negative mass-square. \ See e.g. \ R.G.H.Robertson et al.: Phys.
Rev. Lett. 67 (1991) 957; \ A.Burrows et al.: Phys. Rev. Lett. 68 (1992)
3834; \ Ch.Weinheimer et al.: Phys. Lett. B300 (1993) 210; \ E.Holtzshuh et
al.: Phys. Lett. B287 (1992) 381; \ H.Kawakami et al.: Phys. Lett. B256
(1991) 105, \ and so on. \ See also the reviews or comments by M.Baldo
Ceolin: ``Review of neutrino physics", invited talk at the ``XXIII Int.
Symp. on Multiparticle Dynamics (Aspen, CO; Sept.1993); \ E.W.Otten: Nucl.
Phys. News 5 (1995) 11. \ In very recent papers, J.Ciborowski and 
J.Rembielinski claimed to be able to explain a crucial experimental detail
just by the Superluminal hypothesis (``An explanation of anomalies in the
electron energy spectrum for tritium decay", Preprint submitted to 
HEP97--PA10\#744; 1997); while S.Giani (IT, Cern) claimed that, for explaining
the arrival times of the neutrinos (emitted by the 1987a supernova) detected 
by the Monte Bianco and Kamiokande experiments, Superluminal propagation 
speeds are in order.\\

{\em Third: Galactic ``Mini-Quasars", etc.} \ (Apparent Superluminal
expansions observed inside quasars, some galaxies, and --as discovered very
recently-- in some galactic objects, preliminarily called ``mini-quasars") --
Since 1971 in many quasars (and even a few galaxies) apparent
Superluminal expansions were observed [``Nature", for instance, dedicated to
those observations a couple of its covers]. \ Such seemingly Superluminal
expansions were the consequence of the experimentally measured angular
separation rates, once it was taken into account the (large) distance of the
sources from the Earth. \ From the experimental point of view, it is enough
to quote the book ``Superluminal Radio Sources", ed. by J.A.Zensus and
S.Unwin (Cambridge Univ. Press; Cambridge, UK, 1987), and references therein.

\h The distance of those ``Superluminal sources", however, it is
not well known; or, at least, the (large) distances usually adopted have
been strongly criticized by H.Arp et al., who maintain that quasars are much
nearer objects: so that all the above-mentioned data can no longer be easily
used to infer (apparent) Superluminal motions. \ However, very recently,
GALACTIC objects have been discovered, in which apparent Superluminal
expansions occur; and the distance of galactic objects can be more precisely
determined.  \ From the experimental point of view, see in fact the papers
by \ I.F.Mirabel and L.F.Rodriguez. : ``A superluminal source in the Galaxy",
Nature 371 (1994) 46 [with a Nature's comment, ``A galactic speed record",
by G.Gisler, at page 18 of the same issue]; \ and by \ S.J.Tingay et al. (20
authors): ``Relativistic motion in a nearby bright X-ray source", Nature 374
(1995) 141.

\h From the theoretical point of view, both for quasars and
``mini-quasars", see E.Recami, A.Castellino, G.D.Maccarrone and M.Rodon\`o:
``Considerations about the apparent Superluminal expansions observed in
astrophysics", Nuovo Cimento B93 (1986) 119. \ See also E.Recami: ref.($^{10}$), \
and cf. \ R.Mignani and E.Recami: Gen. Relat. Grav. 5 (1974) 615. \ In
particular, let us recall that a {\em single} Superluminal source of light
would be observed: \ (i) initially, in the phase of ``optic boom" (analogous
to the acoustic ``boom" by an aircraft that travels with constant
super-sonic speed) as an intense, suddenly-appearing source; \ (ii) later
on, as a source which splits into TWO objects receding one from the other
with velocity \ $v > 2c$ \ [see the quoted refs.].\\

{\em Fourth: Superluminal motions in Electrical and Acoustical Engineering
-- The ``X-shaped waves"} -- This fourth sector is perhaps the most important 
one. \ Starting with the pioneering work by H.Bateman, it became
slowly known that all the (homogeneous) wave equations ---in a
general sense: scalar, electromagnetic and spinor--- admit solutions with
subluminal ($v < c\/$) group velocities.($^{36}$)  More recently, also
Superluminal ($V > c\/$) solutions have been constructed for those
homogeneous wave equations, in refs.($^{37}$) and quite independently in
refs.($^{38}$): in some cases just by applying a Superluminal Lorentz
``transformation".($^{10,39}$)  \ It has been also shown that the same happens even
in the case of acoustic waves, with the presence in this case of
``sub-sonic" and ``Super-sonic" solutions.($^{40}$)  \ Particular attention 
has been
called by the circumstance that some of the new solutions are ``undistorted
progressive waves" (namely, represent localized, non-diffractive waves). \ 
One can expect all such solutions to exist, e.g., also for seismic wave 
equations. \ More intriguingly, we might
expect the same to be true in the case of gravitational waves too. 

\h It is interesting to remark that the Super-sonic and
Super-luminal solutions forwarded in refs.($^{41}$) ---some of them
experimentally already realized($^{41}$)--- appear to be (generally speaking)
X-shaped, just as predicted in 1982 by A.O.Barut, G.D.Maccarrone and E.Recami
in ref.($^{42}$); so that now they have been preliminarily called ``X-waves".

\h On this regard, from the theoretical point of view, let us
quote pages 116-117, and pages 59 (fig.19) and 141 (fig.42), in E.Recami:
ref.($^{10}$).  \ Even more, see the abovementioned A.O.Barut, G.D.Maccarrone and
E.Recami: ``On the shape of tachyons", Nuovo Cimento A71 (1982) 509-533;
where ``X-shaped waves" are predicted and discussed. From the quoted papers
it is also clear why the X-shaped waves keeps their form while travelling
(non-diffractive waves): a property that already resulted of high interest
for electrical and acoustical engineering. \ New experimental and theoretical
work is going on (e.g., by F.Fontana et al. at the ``Pirelli Cavi", Milan,
Italy; and by H.E.Hern\'andez F. et al. at the F.E.E.C. of Unicamp, Campinas,
S.P.); let us mention in particular that by P.Saari, H.S\~onajalg et al. at 
Tartu, Estonia [see e.g. Opt. Lett. 22 (1977) 310; Laser Phys. 7 (1977) 32].

\h Very recently, some further important articles
appeared. Let us mention in particular that: \ (i) at Tartu (Estonia) they 
have confirmed the experimentally production of X-shaped 
(Superluminal) light waves, in optics: see P.Saari and K.Reivelt: ``Evidence 
of X-shaped propagation-invariant localized light waves", appeared in 
{\em Phys. Rev. Lett.}, Nov.24, 1997; \ (ii) simultaneously, (non-truncated) 
X-shaped beams with {\em finite} total {\em energy} ---expected to exist on
the basis of ER--- were mathematically constructed by I.Besieris, 
M.Abdel-Rahman, A.Shaarawi and A.Chatzipetros in the work ``Two fundamental 
representations of localized pulse solutions to the scalar wave equation", 
to appear in {\em J. Electromagnetic Waves Appl.} (1998).

\end{document}